\documentstyle[aps,eqsecnum,twocolumn]{revtex}
\begin{document}
\draft
\title{The big bang as a higher-dimensional shock wave}

\author{P. S. Wesson \thanks{Correspondence to P. S. Wesson by mail, phone:
(519)885-1211 ext.2939, fax: (519)746-8115, e-mail:
wesson@astro.uwaterloo.ca}}
\address{Astronomy Group, Department of Physics, University of
  Waterloo, Waterloo, ON, N2L 3G1, Canada}

\author{H. Liu}
\address{Department of Physics, Dalian University of Technology, Dalian,
  116023, P. R. China}

\author{S. S. Seahra}
\address{Astronomy Group, Department of Physics, University of
  Waterloo, Waterloo, ON, N2L 3G1, Canada}

\date{\today}
\maketitle

\begin{abstract}
We give an exact solution of the five-dimensional field equations
which describes a shock wave moving in time and the extra
(Kaluza-Klein) coordinate.  The matter in four-dimensional
spacetime is a cosmology with good physical properties.  The
solution suggests to us that the 4D big bang was a 5D shock wave.
\end{abstract}

\section{Introduction}
The idea that the universe was created from nothing has a very
long history.  Some highlights in the scientific literature
include the argument that the big bang was a transition from an
earlier four-dimensional Minkowski space to a later space with
standard Friedmann-Robertson-Walker properties
\cite{bonn60,wess85}.  It is also possible in principle that the
big bang was a quantum tunneling event from nothing into 4D de
Sitter space \cite{vil82}.  These and other ideas connected with
inflation can be put on a firmer basis if the manifold is extended
from 4D to higher dimensions (for a review see \cite{over97}). For
example, it is well known that any solution of 4D general
relativity can be embedded in a \emph{flat} 10D space.  In what
follows, we will use the minimal extension from 4D (Einstein)
space to 5D (Kaluza-Klein) space to argue that the 4D big bang may
be the signature of a 5D blip or shock wave.

We will draw on recent results in three areas of
higher-dimensional cosmology.
\begin{enumerate}

\item[(a)]
We can take an empty 5D space that is in general curved and derive
from it a matter-filled 4D space \cite{over97,wess92,col94}. That
such models have matter in 4D but are empty in 5D follows from new
work on an old theorem of differential geometry due to Campbell
(1926). Let us consider a solution of the 4D Einstein field
equations $G_{\alpha\beta}=8\pi T_{\alpha\beta}$, where
$G_{\alpha\beta}$ is the Einstein tensor and $T_{\alpha\beta}$ is
the energy-momentum tensor.  (Here and elsewhere we use a choice
of units to set the gravitational constant and the speed of light
equal to unity; lowercase Greek letters run 0,123 and uppercase
Latin letters run 0,123,4.)  Then it can be shown that the 4D
Einstein equations can be locally embedded in the field equations
of 5D Kaluza-Klein theory \emph{without} sources, which are given
in terms of the Ricci tensor by $R_{AB}=0$
\cite{rom96,lid97,wess99}. This is a very powerful theorem.

\item[(b)]
We can take an empty and \emph{flat} 5D space and embed in it
matter-filled \emph{curved} 4D spaces \cite{ab96,liu98,wess94}.
This means that the 4D big bang could be an artifact of a bad
choice of 5D coordinates.

\item[(c)]
We can study \emph{wave-like} solutions of $R_{AB} = 0$
\cite{liu93,liu94,billy96,saj98,saj99}. Some of these have
remarkable physical properties, and indicate that the big bang
could have been a quantum transition from an oscillating to a
growing (inflationary) mode.

\end{enumerate}

In the next section, we will combine results from the above three
areas to derive an exact solution in 5D which has good physical
properties in 4D and implies a significant change in how we can
view the big bang.

\section{A 5D shock wave and the 4D big bang}
We choose coordinates $x^{A}=t,r\theta\phi,l$ with
$d\Omega^{2}\equiv d\theta^{2}+\sin^{2}\theta d\phi^{2}$, as
usual. A wave in the $t/l-$plane should depend on $u\equiv t-l$.
One such solution of $R_{AB}=0$ is given by the following 5D line
element: {\setlength\arraycolsep{2pt}
\begin{eqnarray}\label{metric}
dS^2 & = & b^{2}dt^2-a^{2} \left( dr^{2}+r^{2}d\Omega^{2} \right)
- b^{2}dl^2 \\ a & = & \left( hu \right)^{\frac{1}{2+3\alpha}} \\
b & = & \left( hu \right)^{-\frac{1+3\alpha}{2\left( 2+3\alpha
\right)}}.
\end{eqnarray}}
This may be confirmed either algebraically using the expanded form
of the field equations \cite{wess99} or computationally using a
fast computer package \cite{lake95}.  The class (1)--(3) depends
on two constants, $h$ and $\alpha$.  The first has physical
dimensions of $L^{-1}$ and is related to Hubble's parameter (see
below).  The second is dimensionless and is related to the
properties of matter associated with the solution.  These can be
evaluated using the regular technique, wherein $R_{AB}=0$ is
broken down to $G_{\alpha\beta}=8\pi T_{\alpha\beta}$ with an
induced or effective energy-momentum tensor that depends on the
pressure $p$ and density $\rho$ of a cosmological perfect fluid
\cite{over97}.  There is an associated equation of state, and
after some algebra we find {\setlength\arraycolsep{2pt}
\begin{eqnarray}\label{matter}
p&=&\alpha\rho\\
8\pi\rho&=&\frac{3h^{2}}{\left( 2+3\alpha \right)^{2}}a^{-3\left( 1+
\alpha \right)}.
\end{eqnarray}}We see that $\alpha=0$ corresponds to the late (dust) universe,
and $\alpha=1/3$ corresponds the the early (radiation) universe.

To elucidate the physical properties of the solution, it is
instructive to change from the coordinate time $t$ to the proper
time $T$.  This is defined by $dT=b~dt$, so
\begin{equation}\label{proper}
T = \frac{2}{3} \left( \frac{2+3\alpha}{1+\alpha} \right)
\frac{1}{h} \left( hu \right)^{\frac{3}{2} \left(
\frac{1+\alpha}{2+3\alpha} \right) }.
\end{equation}
The 4D scale factor which determines the dynamics of the model by
(2) and (6) is then
\begin{equation}\label{scale}
a(T) = \left[ \frac{3}{2} \left(
\frac{1+\alpha}{2+3\alpha} \right) hT
\right]^{\frac{2}{3\left(1+\alpha\right)}}.
\end{equation}
For $\alpha=0$, $a(T)\propto T^{2/3}$ as in the standard
(Einstein-de Sitter) dust model.  For $\alpha=1/3$, $a(T)\propto
T^{1/2}$ as in the standard radiation model.  The value of
Hubble's parameter is given by
{\setlength\arraycolsep{2pt}
\begin{eqnarray}\label{hubble}
H \equiv \frac{1}{a}\frac{\partial a}{\partial T} = \frac{1}{a}
\frac{\partial a}{\partial t} \frac{dt}{dT} & = &
\frac{h}{\left(2+3\alpha \right)} \left( hu \right)
^{-\frac{3}{2}\left(\frac{1+\alpha}{2 + 3\alpha}\right)} \\ & = &
\frac{2}{3 \left( 1+\alpha \right) T}.
\end{eqnarray}
For $\alpha=0$ and $1/3$, (9) shows that $H$ has its standard
values in terms of the proper time.  We can also convert the
density (5) from $t$ to $T$ using (6), and find
\begin{equation}\label{density}
8\pi\rho=\frac{4}{3}\frac{1}{\left(1+\alpha\right)^{2}}\frac{1}
{T^{2}}.
\end{equation}
For $\alpha=0$ we have $\rho=1/6\pi T^{2}$, and for $\alpha=1/3$
we have $\rho=3/32\pi T^{2}$, the standard FRW values.  Thus, the
5D solution (1)--(3) contains 4D dynamics and 4D matter that are
the same as in the standard 4D cosmologies for the late and early
universe.

However, while the 5D approach does no violence to the 4D one, it
adds significant insight.  The big bang occurs in proper time at
$T=0$ by (10); but it occurs in coordinate time at $a=0$ or
$u=t-l=0$ by (5) and (2).  Now the field equations $R_{AB}=0$ are
fully covariant, so any choice of coordinates is valid. Therefore,
we can interpret the physically-defined big bang either as a
singularity in 4D or as a hypersurface $t=l$ in 5D.  Both
interpretations are mathematically valid, so the choice is to a
certain extent philosophical.  Our opinion is tipped by a closer
examination of the solution (1)--(3) using a computer package
\cite{lake95}.  It shows that not only is $R_{AB}=0$, but the
Riemann-Christoffel tensor is $R_{ABCD}=0$ also.  This puts the
solution (1)--(3) into the same mathematical class as others in
the literature \cite{ab96,liu98,wess94}).  But this fact also puts
the solution into a new physical class: it is a plane wave or
soliton moving in a \emph{flat} and empty 5D space. [In 5D, the
group of coordinate transformations $x^{A}\rightarrow
x^{A}(x^{B})$ is wider that the 4D group $x^{\alpha}\rightarrow
x^{\alpha}(x^{\beta})$, so $x^{4}$-dependent transformations are
mathematically equivalent in 5D but physically
\emph{non}-equivalent in 4D.  In principle it is possible to find
coordinate transformations between all metrics with $R_{ABCD}=0$,
but in practice the algebraic complexity involved makes the task
presently impossible.]  In other words, we can view the big bang
either as a singularity in 4D, or as a \emph{non}-singular event
in 5D. In the latter interpretation, it is analogous to a 3D shock
wave passing through a 2D surface.

\section{Conclusion}
We have given an exact solution (1)--(3) of the 5D field equations
$R_{AB}=0$ which when reduced to the 4D field equations
$G_{\alpha\beta}=8\pi T_{\alpha\beta}$ describes a cosmology with
good physical properties (4), (5), (10) and good dynamics (7),
(9).  Kaluza-Klein gravity agrees with the classical tests of
relativity in the solar system \cite{kal95,will92}; and the
cosmological solution gives back the same properties as the 4D
Friedmann-Robertson-Walker models with flat space sections, so to
this extent it is agreement with astrophysical data \cite{leo95}.
However, the solution adds the insight that the singular 4D big
bang may be viewed as a non-singular 5D shock wave.

The mere existence of solutions (1)--(3) raises fundamental
questions about observational cosmology.  Is the universe
higher-dimensional?  (This is implied by particle physics, and
what we have done above can clearly be extended to 10D
superstrings and 11D supergravity: see \cite{over97}).  If there
are extra dimensions, then what coordinate system is practical
cosmology using?  (There is no big bang in a geometrical sense in
5D, but there is in a physical sense in 4D because of the choice
of time ; see \cite{wess94}).  It seems to us that these questions
can be answered empirically.  The best way appears to involve the
3 K microwave background radiation.  In the conventional 4D view,
this is thermalized in the big-bang fireball.  In the
higher-dimensional view, some other mechanism must operate, such
as a variation of particle masses that leads to efficient Thomson
scattering \cite{hoy75}.  We need to look into the detailed
physics and decide by observational data which is the best
approach.

\acknowledgements
We thank A. Billyard and J. M. Overduin for comments, and NSERC
for financial support.

\end{document}